\documentclass[fleqn,twoside]{article}%
\usepackage{amsmath}
\usepackage{amsfonts}
\usepackage{amssymb}
\usepackage{graphicx}
\usepackage{caption}
\usepackage{cite}

\def\be{\begin{equation}}
\def\ee{\end{equation}}
\newcommand{\bey}{\begin{eqnarray}}
\newcommand{\eey}{\end{eqnarray}}
\def\bes{\begin{equation}\begin{split}&}
\def\bi{\bibitem}

\title  {Inflation with F(T) Teleparallel Gravity.}
\author{Manas Chakrabortty$^1$, Nayem Sk$^{2}$, Susmita Sanyal$^3$ and Abhik Kumar Sanyal$^4$.\\
~~~~~~~~~~~~~~~\\
$^1$Dept. of Physics, Bankura University, Bankura, India - 722155\\
$^2$Dept. of Physics, Saidabad Manindra Chandra Vidyapith, Murshidabad, India - 742103\\
$^3,^4$Dept. of Physics, Jangipur College, Murshidabad, India - 742213\\}
\begin{document}
\maketitle
\footnote{
Electronic address:\\
$^{1}$manas.chakrabortty001@gmail.com\\
$^2$nayemsk1981@gmail.com\\
$^3$susmitasanyal@yahoo.com\\
\noindent $^4$sanyal\_ ak@yahoo.com\\}
\begin{abstract}

\noindent
We study early universe with a particular form of F(T) Telleparallel gravity theory, in which inflation is driven by a scalar field. To ensure slow rollover, two different potentials are chosen in a manner, such that they remain almost flat for large initial value of the scalar field. Inflationary parameters show wonderful fit with the presently available Planck's data set. The energy scale of inflation is sub-Planckian and graceful exit from inflation is also administered. The chosen form of F(T) administers late-time cosmic acceleration too. In the process, unification of the early inflation with late-time acceleration is ensured. Unfortunately, a decelerated radiation dominated era is only possible with a different form of (quartic) potential, which being devoid of a flat section does not admit slow rollover.\\
\end{abstract}
\maketitle

\section{Introduction}

SN1a supernovae observations \cite{1.1, 1.2, 1.3} suggest that the universe is currently undergoing an accelerating phase of expansion. Such an uncanny late-time cosmic evolution, is also supported by some other observational evidences like WMAP \cite{1.4}, X-ray \cite{1.5}, LSS \cite{1.6} and SDSS \cite{1.7}. Clearly, standard FLRW model of cosmology, which ensures only decelerating phase of expansion throughout the evolution, does not explain the late time cosmic acceleration. It remains a challenge to the cosmologists for over a couple of decades, to justify such accelerated expansion of the universe. The fundamental requirement of such accelerated expansion is the presence of negative pressure, while thermodynamic pressure due to baryonic and non-baryonic (CDM) matters are positive definite $p \ge 0$. The best known candidate with a negative pressure is the cosmological constant ($\Lambda$), which can unanimously resolve the puzzle, and is dubbed as the $\Lambda$CDM model. However, cosmological constant, which essentially is the vacuum energy density of the universe calculated by the field theorists, is some 120 order of magnitude larger than that required to explain late time cosmic acceleration. Therefore, dynamical models were introduced, by invoking one or more exotic field(s). These are called dark energy (DE) models, since neither cosmological constant nor such fields interact with anything other than gravity. Till date, other than the Higgs, which is not responsible for such late-stage accelerated expansion, no scalar has been detected\footnote{There is a recent indication for direct detection of dark energy in the world's most sensitive WIMP detector XENON1T, located thousands of feet underneath the Monte Gran Sasso, Italy. Last year it puzzled the scientists by reporting an excess of about 53 recoil electrons. Using chameleon screening technique, a group of scientists has claimed that these excess electrons might be an outcome of interaction with dark energy (S. Vagnozzi et al, (2021) Phys. Rev. D 104, 063023).}. An alternative to dark energy models has therefore been invoked, which requires to modify General theory of Relativity (GTR) by introducing higher order curvature invariant terms. These models are known as ``modified theory of gravity". A host of such alternatives to dark energy models viz., ``$F(R)$ gravity" \cite{1.8,1.9,1.10,1.11,1.12,1.13,1.14,1.15,1.16}, ``$F(G)$ or Gauss Bonnet gravity" \cite{1.17,1.18,1.19}, ``F(R) Ho$\check{\mathrm{r}}$ava-Lifshitz gravity" \cite{1.20,1.21,1.22,1.23},``Lovelock gravity" \cite{1.24,1.25,1.26,1.27}, their combinations \cite{1.28,1.29,1.30}, and even more, appear in the literature. In the present manuscript, our concern is with Teleparallel gravity theory, which has drawn lot of attention in recent years. \\

In analogy to the $F(R)$ theory of gravity, recently a new modified theory of gravity, namely the so-called $F(T)$ theory of gravity, also dubbed as ``Gravity with torsion" has been proposed to explain current accelerated expansion without invoking dark energy \cite{1.31,1.32,1.33}. This is essentially a generalized version of the so-called `Teleparallel gravity'. Teleparallelism was first attempted by Einstein \cite{1.34} to base a unified theory of electromagnetism and gravity, on the mathematical structure of distant parallelism. However, the scheme failed. New Teleparallel gravity theory is a theory of gravitation based on Weitzenb\"ock spacetime \cite{1.35}, and attributes gravitation to the torsion tensor formed out of the parallel vector fields. It is important to mention that, there is no foundational reason to consider torsion-less space-time, other than its simplicity. However, it is required to test its advantage over GTR, which is our present concern. For a comprehensive review of $F(T)$-gravity and its cosmological implications, see \cite{1.35a} and references therein. Let us begin with a brief review of the modified Teleparallel gravity theory. The action of $F(T)$ gravity is given by,

\be\label{1.1}\mathbb{ A} = \int  d^4 x  \mid e \mid  F(T )+ S_m ,\ee
where $|e|$ = det $e^{i}_{\mu}=\sqrt {-g}$, and the units has been chosen so that $c = 16 \pi G = 1$. Teleparallelism uses a vierbein field  $ \mathbf{e_{i}} (x^{\mu}),~ i = \{0, 1, 2, 3\}$, as dynamical object, which is an orthonormal basis for the tangent space at each point $x^{\mu}$ of the manifold: $\mathbf{e_{i}}.\mathbf{e_{j}}={\eta}_{ij}$, where ${\eta}_{ij}$ = diag(-1,1,1,1). Each vector $\mathbf{e_{i}}$ can be described by its components $e^{\mu}_{i}, ~\mu = \{0, 1, 2, 3\}$ in a coordinate basis; i.e. $\mathbf{e_{i}}=e^{\mu}_{i}\partial\mu$. Here, the Latin indices refer to the tangent space, while Greek indices label coordinates on the manifold. The metric tensor is obtained from the dual vierbein as $ g_{\mu\nu}(x)=\eta_{ij}e^{i}_{\mu}(x) e^{j}_{\nu}(x)$. In contrast to the GTR, which uses the torsion-less Levi-Civita connection, Teleparallelism uses the curvature-less Weitzenb$\ddot{\mathrm{o}}$ck connection \cite{1.35}, whose non-null torsion is

\be\label{1.2} T^{\lambda}_{\mu\nu} \equiv e^{\lambda}_{i}[\partial_{\mu}e^{i}_{\nu}-\partial_{\nu}e^{i}_{\mu}].\ee
The above tensor encompasses all the information regarding the gravitational field. The Teleparallel equivalent of General Theory of Relativity (TEGR) Lagrangian is built with the torsion \eqref{1.2}, and its dynamical equations for the vierbein lead to Einstein equations for the metric. The Teleparallel Lagrangian is given by \cite{1.36,1.37,1.38},

\be\label{1.3} L_T = {S_{\rho}}^{\mu\nu} {T^{\rho}}_{\mu\nu},\ee
where,
\be\label{1.4} {S_{\rho}}^{\mu\nu} =  \frac{1}{2}[{K^{\mu\nu}}_{\rho}+{\delta}^{\mu}_{\rho}{T^{\theta\nu}}_ {\theta}-{\delta}^{\nu}_{\rho}{T^{\theta\mu}}_{\theta}],\ee
while $K^{\alpha\beta}_{\rho}$ is the contorsion tensor given by,

\be\label{1.5} {K^{\mu\nu}}_{ \rho} =  -\frac{1}{2}[{T^{\mu\nu}}_{\rho}-{T^{\nu\mu}}_ {\rho}-{T_{\rho}}^{\mu\nu}],\ee
which equals the difference between Weitzenb$\ddot{\mathrm{o}}$ck and the Levi-Civita connections.\\

$F(T)$  Teleparallel theory of gravity was primarily introduced to drive inflation by Ferraro and Fiorini \cite{1.39,1.40}. Later, Bengochea and Ferraro \cite{1.41}, Linder \cite{1.42}, and also Myrzakulov \cite{1.42a} proposed to use $F(T)$ Teleparallel theory of gravity to drive the current accelerated expansion of our universe, as an alternative to dark energy. The theory has thereafter been studied extensively over last decade. For example, Hamiltonian constraint analysis in $F(T)$ Teleparallel gravity has been studied \cite{1.43,1.44,1.45,1.46}. Further, constraints on $F(T)$ Teleparallel gravity \cite{1.47,1.48} by latest observational data-sets, analysis of the dynamical behaviour \cite{1.49} and the cosmic large scale structure \cite{1.50}, relativistic neutron star \cite{1.51}, matter bounce \cite{1.52} and perturbations \cite{1.53,1.54} etc. have also been explored. Additionally, in $F(T)$ Teleparallel gravity framework, static spherical symmetry solutions \cite{1.55,1.56,1.57}, validity of Birkhoff's theorem \cite{1.58,1.59,1.60}, Solar system tests \cite{1.61,1.62,1.63}, black hole solutions \cite{1.64,1.65,1.66,1.67}, wormhole solutions \cite{1.68,1.69,1.70} and the equation-of-state (EOS) parameter crossing the phantom divide line, have also been explored \cite{1.71}. In the theoretical aspect, the Lorentz invariance and conformal invariance of the $F(T)$ Teleparallel theory are also investigated \cite{1.72,1.72a,1.73}, and many interesting results emerged in the process. Finally, Inflation in the context of $F(T)$ gravity has also been explored extensively \cite{Inf1,Inf2,Inf3,Inf4,Inf5,Inf6,Inf7,Inf8,Inf9,Inf10,Inf11,Inf12,Inf13,Inf14,Inf15,Inf16,Inf17} and in some works inflationary parameters were matched with experimentally observed data. However, most of these tests were made prior to the currently released data from Planck collaborations \cite{Planck1, Planck2}. Our present concern is primarily to test Inflation and match the inflationary parameters with the latest release data from Planck's collaborations \cite{Planck1, Planck2}, and also to check if it is compatible with later epoch of cosmic evolution in the matter dominated era. \\

Latest released data from Planck collaborations \cite{Planck1, Planck2} has dramatically tightened the constraints on tensor to scalar ratio $(r)$, the slope of the primordial scalar power spectrum, conventionally parameterized by the power-law index $(n_s)$, and the tensor spectral index $(n_t)$. The combined data (TT, TE, EE + lowE + lensing + BK15 + BAO + Bicep2) has constrained $r$ to $r < 0.058$, sets the range on $n_s$ in the limit $n_s = 0.9668 \pm 0.0037$, with a very small tensor spectral index $n_t$, fixed by the single field slow-roll self consistency condition. Now to test inflation, a specific form of $F(T)$ is required. Noether symmetry analysis have been extensively performed in this respect \cite{1.74, 1.75, 1.76, 1.77, 1.78, 1.79}. However, through a private communication it is learnt that none of the forms of $F(T)$ associated with their conserved currents so obtained, satisfy the energy constraint equation of Einstein \cite{NM}. On the contrary, the outcome of reconstruction program are ($F(T)\propto \sqrt T$ in $\Lambda$CDM model, $F(T)\propto T$ in the pressure-less dust model and $F(T)\propto (T^2+6\beta T-3{\beta}^2)$) for stiff fluid \cite{1.42a}, out of which the last form appears to be quite reasonable. In fact several authors followed reconstruction mechanism to find $F(T) = \alpha T + \beta T^2$ (apart from some complicated forms), some of which we shall discuss shortly. It is noteworthy that this particular form is the simplest modification, and has also been found through Noether symmetry analysis, although as mentioned, the associated conserved current is not compatible with the field equation. We therefore choose this particular form of $F(T) = \alpha T + \beta T^2$, by hand for our purpose.\\

Let us briefly review the current status of theoretical study on inflation in Teleparallel $F(T)$ theory. Authors \cite{Inf1} studied inflation with non-minimal gravitational coupling of the torsion scalar and the electro-magnetic field, which breaks the conformal invariance. They found generation of large-scale magnetic field with field strength of the order of $10^{-9}~G$, on $1~ Mpc$ scale. This is sufficient to account for the large-scale magnetic fields observed in clusters of galaxies only through the adiabatic compression, during the construction of the large scale structure of the universe. This does not rewuire the dynamo amplification mechanism. In \cite{Inf2} authors studied trace-anomaly driven inflation in $T^2$ Teleparallel gravity theory. In particular, they demonstrated that the de Sitter inflation can be realized in $T^2$ gravity, and graceful exit is realized. In \cite{Inf3}, warm intermediate inflation $a \propto a_0 e^{At^f}, A > 0,~ 0 < f < 1$ is studied with $F(T) = T +\alpha \sqrt{-T}$, in the presence of a minimally coupled scalar field. However, the plots depict that either $n_s$ or $r$ lie within the current observational limit, but not the two together. For example, if $n_s \approx 0.96$, then $r > 0.2$, in the first case and $r > 0.1$ in the second. On the contrary, if $r$ is kept within experimental limit, $r < 0.06$ (say), then $n_s ~ 0.5$ in the first case, and $n_s ~ 0.1$ in the second, which are awfully bad. Nonetheless, the model admits graceful exit from inflation. In \cite{Inf4}, authors chose $F(T) = T + \beta T^2$, minimally coupled to a scalar field and reconstructed the scalar potential as: $V(\phi)=A + B e^{-\sqrt{1\over 2n}\phi}$. Although $n_s= 0.9691$ lie within experimental limit, $r=0.248$ is far from recently established constraint $r < 0.06$ \cite{Planck1, Planck2}. Further, the model does not admit slow roll. Additionally, it is not clear how the authors obtained $a(t) \propto t^{2\over 3(1 + \omega)}$, (where $\omega$ is the equation of state parameter) which matches with Friedmann-solution in matter dominated (radiation and pressure-less dust) eras. Next, the authors in \cite{Inf5} investigated perfect fluid description of slow-roll parameters, and reconstructed $F(T)$. For quite a complicated power-law form of $F(T)$, the slow-roll parameters have been found to be consistent with the Planck data. However, consideration of perfect fluid in the very early universe is questionable. The authors in \cite{Inf6} studied inflation considering power law $F(T) = T^n$, in the presence of a canonical scalar field, while the scale factor has been chosen to admit power law inflation and intermediate inflation. The potential required is $V(\phi) = \phi^m$. Self-interacting quartic potential is found to be consistent with Planck (2015) data. The unanswered question is: how the scalar slow rolls with quartic potential? The authors of \cite{Inf7} derived a particular class of $F(T)$ model, which is identified with flat-like universe. With a minimally coupled scalar field, double slow roll inflation is realized, and a quasi-inverse power law inflation has been found to fit with the observed data, which admits graceful exit as well. Authors of \cite{Inf8} reconstructed a complicated form of $F(T)$ and investigated bounce inflation with a canonical scalar field. Although graceful exit is admissible, they found $r = 0.00156$ and the scalar tilt $n_s = 0.997$, exhibiting nearly scale invariant power spectrum, which is ruled out by all observations. Following reconstruction mechanism, authors of \cite{Inf9} found a highly complicated form of $F(T)$, which reduces to $F(T) = c_1 \sqrt{-T} + c_2$ in vacuum (note that such a form of $F(T)$ makes the action singular). Inflation is studied in the context of unimodular $F(T)$ gravity. Although, graceful exit is admissible, the spectral index $n_s \approx 0.98$ however, exceeds current experimental limit. In \cite{Inf10}, assuming intermediate inflation $a(t) = a_0 e^{At^n}$, with $A > 0,~ 0 < n < 1$, for $F(T) = T + f(T)$, authors found $f(T) = c_1 T^m – {T\over 2(1-m)}, m = {An\over 2}$ in view of the evolution of perturbation. Although the spectral index $n_s = 0.9644 \pm 0.0049$ lies very much within the experimental limit, the tensor to scalar ratio could not be found. Logamediate inflation $a(t) = a_0 e^{A(ln~t)^\lambda}$, with $A > 0,~ 0 < \lambda < 1$ was studied in \cite{Inf11}, taking into account $F(T) = T_0 \left({T\over T_0}\right)^n$. Quite a nice fit with observed data was found with Planck (2015) TT, TE, EE + lowP data, for $n = 2, \lambda = 8$. Under the choice $F(T) = \alpha T + \beta T^2$ authors in \cite{Inf12} studied constant roll inflation with a minimally coupled inflaton field. Perturbative analysis was also carried out successfully. However, although the authors found $n_s  = 0.96$, hiwever $r_{min}= 0.08$ crosses experimental limit. Considering non-minimal coupling with a tachyonic field, authors of \cite{Inf13} found $N = 58$, $n_s = 0.956$ and $r = 0.0061$. Clearly $n_s$ lie much below the experimental data. Authors of \cite{Inf14} considered a minimally coupled scalar field to explore power law inflation, the scalar field being responsible to reheat after inflation. The authors obtain an expression for the reheating temperature in terms of the CMB temperature, the spectral index, the power spectrum and the parameters of the model. In another work \cite{Inf15}, taking into account a canonical scalar field, non-minimally coupled to the torsion with a Galileon-type self-interaction, exhaustive study on different slow roll inflationary scenario on generalized scalar-torsion gravity, and excellent agreement with Planck’s data were established. Nonetheless, since the authors considered $F(T) = T$, so it is the K-essence model of GTR, in disguise. The authors of \cite{Inf16} reconstructed $F(T) = T + T^2 –c$  near the type IV finite-time singularity in the Jordan frame. Considering a specific form of Hubble rate, they confirm theoretical $F(T)$ description based on slow-roll parameters ($n_s = 0.966, r < 0.07$) with Planck and BICEP2/Keck-Array data. Graceful exit is also admissible in the model. Further, the above form of $F(T)$ has been claimed to unify the early and late-time evolution of the universe. Last but not the least, the authors \cite{Inf17} show that a Teleparallel theory with a non-minimally coupled Higgs scalar field has no linear scalar perturbations, and therefore cannot give successful inflation, unless the non-minimal coupling functions satisfy a particular relation. On the contrary, if the relation is satisfied, Higgs inflation can give rise to an arbitrarily large tensor-to-scalar ratio $r$. The results also apply to $F(T)$ theories, as they are scalar-tensor theories written in different field coordinates.\\

In a nut-shell, except for the recent work \cite{Inf16}, in which a specific form of Hubble parameter is chosen, none other fits perfectly with the current released data set \cite{Planck1, Planck2}. However, what happens in the radiation dominated era has not been exhibited in \cite{Inf16}. In this respect, our motivation is to unify a Hubble parameter driven late-time acceleration and a scalar field driven slow roll inflation. Further, we study cosmological evolution in the radiation dominated era. It is noticeable that, in almost all of the above mentioned works, inflation is driven by a scalar field, while a form of $F(T)$ has been found under reconstruction programme, or sometimes following perturbative analysis. Note that, the same field cannot be responsible for accelerating the universe both in the early and the late stage of cosmological evolution. This means, if a particular form of $F(T)$ is found responsible for late-time cosmic acceleration via reconstruction program (say), then a different field (curvature or a scalar) is required for early inflation. In the context of purely geometric $F(R)$ theory of gravity, it was therefore suggested to consider an action in the form $A = \sqrt{-g} d^4x[\alpha R + \beta R^2 + \gamma R^{-n}]$ \cite{1.80} or $A = \sqrt{-g} d^4x[\alpha R + \beta R^2 + \gamma R^{2\over 3}]$ \cite{1.81}, so that $R^2$ may be responsible for inflation in the early universe, $R$ in the middle to ensure Friedmann-like matter dominated eras, and $R^{-n}$ or $R^{2\over 3}$ at the late stage of cosmic evolution, to ensure current accelerated universe. Thus, direct coupling of a scalar field to the torsion scalar in Teleparallel gravity has been studied extensively over last decade (references are available in \cite{Inf17}). The simplest of these models, considered by several authors, is $F(T) = \alpha T + \beta T^2$, which can produce acceleration at the late-stage of cosmic evolution.\\

In view of the above discussion, we also shall consider a minimally coupled scalar field (inflaton) to drive inflation in the very early universe, and try to fit inflationary parameters with the currently released data set \cite{Planck1, Planck2}. In the following section we cast the field equations in the background of spatially flat Robertson-Walker metric. Since vacuum de-Sitter solution is realized for arbitrary form of $F(T)$, we choose a particular form: $F(T) = \alpha T + \beta T^2$, by hand, and write down the field equations in the presence of a minimally coupled scalar field. Next, we show that indeed such a form of $F(T)$ envisages cosmic acceleration in the current pressure-less dust era. In section 3, we study inflation, being driven by the scalar field, under slow roll assumption. In section 4, we try to find an analytical solution from the field equations in the radiation dominated era. Finally, we conclude in section 5.

\section{Field equations and cosmological solutions:}

In the spatially flat Robertson-Walker (RW) space-time,
\be\label{2.1} {ds}^2 = - {dt}^2 + {a^2(t)}\big[dr^2 + r^2 \big(d\theta ^2 + \sin^2{\theta}~d\phi^2\big)\big],\ee
the components of vierbein field are expressed in terms of the cosmological scale factor $a(t)$ as  ,

\be\label{2.2}e^{i}_{\mu}= \mathrm{diag}(1,a(t),a(t),a(t)).\ee
If matter couples to the metric in the standard form, then the variation of the action with respect to the vierbein leads to:

\be\label{2.3} e^{-1}\partial{\mu}(e S^{\mu\nu}_{i})F(T)_{,T}-e^{\lambda}_{i}T^{\rho}_{\mu\lambda}S^{\nu\mu}_{\rho}F(T)_{,T}+S^{\mu\nu}_{i}\partial_{}\mu(F(T))F(T)_{,TT}+
\frac{1}{4}e^{\nu}_{i}F(T)=\frac{1}{4}e^{\rho}_{i} \mathbb{T}^{\nu}_{\rho},\ee
where suffix $T$ denotes differentiation with respect to $T$, $S^{\mu\nu}_{i}=e^{\rho}_{i}S^{\mu\nu}_{\rho}$ and $\mathbb{T}_{\mu\nu}$ is the matter energy-momentum tensor. Now, for the spatially flat Robertson-Walker (RW) metric under consideration, equations \eqref{1.2}, \eqref{1.4} and \eqref{1.5} lead to

\be\label{2.4} T = {S}^{\rho\mu\nu} {T}_{\rho\mu\nu}=6\left(-\frac {\dot a^2}{a^2}\right)=-6\frac{\dot a^2}{a^2}=-6H^2,\ee
$H$ being the Hubble parameter. Substituting the vierbein \eqref{2.2} in \eqref{2.3} for $i = 0 = \nu$, one obtains the following  field equation, in the presence of a barotropic fluid and minimally coupled scalar field,

\be\label{2.5} F +12H^2F_{,T} =\rho + \rho_\phi = \rho + {1\over 2} \dot\phi^2 + V(\phi),\ee
\be\label{2.6} 48H^2\dot{H}F_{,TT} - 4(\dot{H}+3H^2)F_{,T}-F = p + p_\phi = p + {1\over 2} \dot\phi^2 - V(\phi).\ee
Equation \eqref{2.5} is essentially the $(^0_0)$ component of Einstein's equation, while \eqref{2.6} is the field equation for $i = 1 = \nu $. Note that \eqref{2.6} is the only independent $(^i_i)$ component of Einstein's equations. In the above, $p$ and $\rho$ stand for thermodynamic pressure and density respectively for the barotropic fluid under consideration. Thus, equations \eqref{2.5} and \eqref{2.6} constitute Einstein's cosmological field equations for $F(T)$ gravity in spatially flat RW metric \eqref{2.1}, under the condition \eqref{2.4}. Let us mention that the above field equations may also be found using Lagrange multiplier technique, as well as from scalar-tensor equivalent forms \cite{STe1, STe2}, as in the case of $F(R)$ theory of gravity. It is trivial to check that GTR may be recovered for $F(T) \propto T$, apart from a boundary term.\\

Now, the field equations in pure $F(T)$ gravity (in the absence of the scalar field) are,
\be \label{2.20} F + 4(\dot{H}+3H^2)F_{,T} - 48H^2\dot{H}F_{,TT} = 0,\ee
\be \label{2.21} F +12H^2F_{,T} = 0.\ee
The above set of equations may be combined to yield,

\be\label{2.22} \dot H(12 H^2 F_{,TT} - F_{,T}) = 0.\ee
Hence, either the field equations admit a solution in the form $F(T) = F_0\sqrt T =i F_0 \sqrt{6}\left({\dot a\over a}\right)$, ($F_0$ being the constant of integration) which is clearly meaningless; or a de-Sitter solution $(\dot H = 0,~ a = a_0 e^{\lambda t})$ for arbitrary form of $F(T)$. As mentioned in the introduction, we associate a scalar field $(\phi)$ to drive inflation, and choose a form of $F(T)$ as,

\be\label{F} F(T) = \alpha T + \beta T^2,\ee
say, where the dimension of $[\alpha] = \big[M_P^2\big]$ and the constant parameter $\beta$ is dimensionless. Note that the sound speed for such a form of $F(T)$ is:

\be c_s^2 = \frac{F_H}{HF_{HH}} =\frac{F_{,T}}{F_{,T}-12H^2 F_{,TT}} = \frac{72\beta H^2-\alpha}{216 \beta H^2-\alpha},\ee
and so $0< c_s < 1$ is always ensured, provided $\beta > 0$. Now, the field equations in the presence of a barotropic fluid are:

\be\label{2.33a} \alpha(2\dot H + 3H^2) - 18\beta H^2(4\dot H + 3H^2) = -{1\over 2}\left[p + {1\over 2}\dot\phi^2 -\ V(\phi)\right],\ee
\be\label{2.33b} \alpha H^2 - 18\beta H^4 = {1\over 6}\left[\rho + {1\over 2}\dot\phi^2 +\ V(\phi)\right],\ee
\be\label{2.33c} \ddot \phi + 3H\dot \phi + V'(\phi) = 0.\ee
In the above, prime denotes derivative with respect to $\phi$. For further clarification regarding the above choice, we find the effective state parameter as,

\be \label{2.33d} \omega_e = \frac{p + {1\over 2}\dot\phi^2 - V(\phi) - 108\beta H^2 (4\dot H + 3H^2)}{\rho + {1\over 2}\dot\phi^2 + V(\phi) + 108\beta H^4}.\ee
In the pressure-less dust era $(p = 0)$, the effective state parameter therefore may remain positive initially, depicting a decelerated expansion. However, at a latter epoch, it turns out to be negative and initiate accelerating expansion. The presence of $H^4$ term with a negative sign in the numerator, might also allow the model to cross the phantom divide line $\omega_e < -1$. This is true even in the absence of the scalar field, if it decays completely (say) in the process of producing particles, while oscillating at the end of inflationary epoch. Clearly, the same $H^4$ term cannot be responsible for driving late-time cosmic acceleration and early inflation simultaneously. Therefore, we need a scalar field to drive inflation at the very early stage of the cosmic evolution.

\section{Inflation under slow-roll approximation:}

If a single scalar field drives inflation, then the standard slow-roll approximations $|\ddot \phi| \ll 3H\dot \phi$ and $\dot\phi^2 \ll V(\phi)$ hold. As a result, we can express \eqref{2.33b} and \eqref{2.33c} respectively as:

\be \label{2.34a}  \gamma H^4-6\alpha H^2 +V(\phi) = 0,\ee
\be \label{2.34b}  3H\dot\phi + V'(\phi) = 0,\ee
in the vacuum era $(p = 0 = \rho)$. In the above, we have redefined $108\beta =\gamma$. Let us now solve $H^2$ from \eqref{2.34a} to obtain,

\be\label{2.34c} H^2 = \frac{3\alpha - \sqrt{9\alpha^2 - \gamma V(\phi)}}{\gamma},\ee
where we have chosen negative sign to ensure the first slow-roll parameter $\epsilon > 0$. Next, we compute the number of e-folds $(N)$as,.

\be \label{2.34d} N = \int_{t_i}^{t_f} H dt = {3\over \gamma}\int_{\phi_f}^{\phi_i} \frac{3\alpha-\sqrt{9\alpha^2-\gamma V}}{V'}d\phi.\ee
Further, in view of equation \eqref{2.34a} we can compute ${\dot H\over H^2} = -{V'^2\over 12H^4[\sqrt{9\alpha^2-\gamma V}]}$, and as a result  the slow roll parameters $\epsilon$ and $\eta$ may be computed in the following manner

\be\label{2.34e} \epsilon = -{\dot H\over H^2} = {\gamma^2V'^2\over 12[\sqrt{9\alpha^2-\gamma V}]\left[3\alpha-\sqrt{9\alpha^2-\gamma V}\right]^2}, ~~\mathrm{and,~~}\eta = 2\alpha\left(V''\over V\right),\ee
where, $\epsilon \ll 1$, and $|n_s| \ll 1$. The tensor to scalar ratio $(r)$, the primordial spectral index of scalar perturbation $(n_s)$, and the tensor spectral index $(n_t)$ may now be expressed as,

\be \label{2.34f} r = 16\epsilon,~~~~~n_s = 1-6\epsilon + 2\eta,~~~~~ n_t = -{r\over 8} = -2\epsilon.\ee
As mentioned in the introduction, the recent released combined data (TT, TE, EE + lowE + lensing + BK15 + BAO + Bicep2) have constrained the inflationary parameters as: $r < 0.058$, $n_s = 0.9668 \pm 0.0037$ and a very small tensor spectral index $n_t$ \cite{Planck1, Planck2}, while number of e-folding preferably should lie within the range $45 < N < 65$, to solve the horizon and the flatness problems. Before choosing a form of potential, let us mention that for the single field inflationary model under present consideration, these constraints imply $V'' < 0$. Since otherwise i.e. for $V'' > 0$, $\eta > 0$, and hence any attempt to keep $n_s < 0.97$, makes $r > 0.08$. Next, we should choose $\phi_i$ in such a manner that $\phi_f \approx 1 M_P$, as inflation ends (i.e. $\epsilon = 1$). The reason being, a graceful exit from inflation requires $\phi$ should oscillate as it becomes less than Planck's mass. Let us make two choices of the potential in the following form,

\be \label{Vphi}V(\phi) = V_0 - {V_1\over \phi},~~ V(\phi) = V_0 - V_1e^{-b\phi},\ee
so that the potential remains almost flat as $\phi$ is large, and slow roll is admissible. From dimensional analysis, it is clear that since $[H] = [M_P]$, so $[\alpha] = [M_P^2],~[\phi]=[M_P]$ and $\gamma$ is dimensionless.\\

\subsection{Case-I:}
Here we consider the potential in the form: $V(\phi) = V_0 - {V_1\over \phi},~~ \mathrm{so~that}~~~ V' = {V_1\over \phi^2}, ~~\mathrm{and},~~ V'' = -{2V_1\over\phi^3}$. We take $\phi_i = 4 M_P,~2\alpha = 1 M_p^2,~{V_0\over V_1} = 2 M_P^{-1}, ~ \gamma V_1 = 1 M_P^5$, to compute slow roll parameters as,

\be \eta = -{2\over \phi_i^2\left({V_0\over V_1}\phi_i-1\right)} \approx -0.017.\ee
\be \epsilon = \frac{\gamma^2V_1^2}{12 \phi_i^4\sqrt{2.25-{\gamma V_1\left({V_0\over V_1} -{1\over\phi_i}\right)}}\left[1.5-\sqrt{2.25-{\gamma V_1\left({V_0\over V_1} -{1\over\phi_i}\right)}}\right]^2} = 0.000732258.\ee
Therefore,
\be r = 16 \epsilon = 0.0117,~~~ n_s = 1 - 6\epsilon + 2\eta = 0.9616,~~~\mathrm{and}~~~n_t = -2\epsilon = -0.00146.\ee
The slow roll over terminates $(\epsilon = 1)$ at $\phi_f = 0.9 M_P$. The number of e-folding $N$, may now be found using \eqref{2.34d} as,

\be\begin{split} N =& {9\alpha\over \gamma}\int_{\phi_f}^{\phi_i} {d\phi\over V'} - {3\over \gamma}\int_{\phi_f}^{\phi_i}{\sqrt{9\alpha^2-\gamma V}\over V'}d\phi = {4.5\over \gamma V_1}\int_{\phi_f}^{\phi_i}\phi^2 d\phi -{3\over \gamma V_1}\int_{\phi_f}^{\phi_i}\sqrt{2.25-\gamma V_1\left({V_0\over V_1}-{1\over \phi}\right)}\phi^2 d\phi\\&
= 4.5{\phi^3\over 3}\Big|_{0.92}^{4}- 3\int_{0.92}^{4}\left(\sqrt{2.25-2+{1\over \phi}}\right)\phi^2 d\phi = 94.83-1.5\int_{.92}^{4}\sqrt{\phi^4+4\phi^3}d\phi\\&
94.83-1.5\times 32.749 = 46.\end{split}\ee
Clearly, this is an excellent fit with the currently available Planck's data set \cite{Planck1, Planck2}. In Table-1, we organize a data set for the inflationary parameters, varying $\phi_i$ between $4 M_P\le \phi_i \le 4.35 M_P$, so that the spectral index of scalar perturbation lie within experimental limit. Note that slow rollover terminates $(\epsilon = 1)$ at $\phi_f =0.9 M_P$ in every case. Further, the number of e-folds $(46 \le N \le 60)$ found, is sufficient to solve the horizon and flatness problems.\\

\begin{figure}
\begin{center}
   \begin{minipage}[h]{0.47\textwidth}
      \centering
      \begin{tabular}{|c|c|c|c|}
      \hline\hline
    $\phi_i$ in $M_P$ & $n_s$ & $r$ & $N$ \\
    \hline
    4.00 & 0.9598 & 0.0117 & 46  \\
    \hline
    4.05 & 0.9614 & 0.0111 & 48  \\
    \hline
    4.10 & 0.9630 & 0.0105 & 50  \\
    \hline
    4.15 & 0.9644 & 0.0100 & 52 \\
    \hline
    4.20 & 0.9657 & 0.0095 & 54  \\
    \hline
    4.25 & 0.9670 & 0.0091 & 56  \\
    \hline
    4.30 & 0.9682 & 0.0086 & 58  \\
    \hline
    4.35 & 0.9694 & 0.0082 & 60  \\
    \hline\hline
    \end{tabular}
      \captionof{table}{Data set for the inflationary parameters under the choice $F(T) =\alpha T + \beta T^2$ and $V = V_0 -{V_1\over \phi}$, where $\alpha = {M_P^2\over 2},~{V_0\over V_1} = 2 M_P^{-1}$, $108\beta V_1 = \gamma V_1 = 1 M_P^5$. Slow rollover ends at $\phi_f = 0.9 M_P$.}
      \label{tab:table1}
   \end{minipage}%
 \end{center}
\end{figure}

Let us now find the energy scale of inflation in view of the relation \eqref{2.34c}. For this purpose, we consider the last data set of the above table-1, associated with $N = 60$, for which $\phi_i = 4.35 M_P$. As a result we find

\be \label{scale} {H_*}^2 = \frac{1.5 -\sqrt{2.25-\gamma V_1\left({V_0\over V_1}-{1\over \phi_i}\right)}}{\gamma} = {0.807\over \gamma}, ~~\mathrm{or} ~~H_* = \sqrt {0.807\over \gamma},\ee
where, we have substituted $\alpha = {1\over 2} M_P^2$, ${V_0\over V_1} = 2 M_P^{-1}$ and $\gamma V_1 = 1 {M^5_P}$. Note that $\gamma$ is still arbitrary. Now, the energy scale of inflation in a single scalar field model corresponding to GTR, is given by the expression \cite{83} as,

\be \label{sf} H_* = 8\times 10^{13}\sqrt{r\over 0.2}~GeV = 1.62\times 10^{13}~GeV = 6.6\times 10^{-6} M_P.\ee
In the above, we have used the value of the tensor to scalar ratio $r = 0.0082$ from the last data set of table-1. Thus in order that the scale of inflation \eqref{scale} matches with the single field scale of inflation \eqref{sf}, $\gamma$ has to be of the order, $\gamma \approx 1.8\times 10^{10}$, i.e. $\beta \approx 10^8$. Further, since individually each term has to be of the same order of magnitude, one can compare the two terms on the left hand side of \eqref{2.34a} to find that:

\be 6\alpha H_*^2 \approx 1.29 \times 10^{-10} M_P^4,~~ \mathrm{while}~~108\beta H_*^4 = \gamma H_*^4 \approx 10^{-10},\ee
provided $\gamma \approx 10^{10}$. It is important to note that, we essentially have to fix only two parameters $\gamma V_1$ and $V_0\over V_1$, for a viable inflationary model. However, in view of the above consideration (sub-Planckian scale of inflation), the coupling parameter $\gamma = 108\beta$ also is fixed, which has been found to match with the value of $\gamma$ in view of equation \eqref{2.34a}. This result is of-course encouraging. In the same manner, one can also guess the value of $V_1$ and hence $V_0$. Since the potential $V(\phi)$ appearing in \eqref{2.34a} has also to be of the same order of magnitude, so

\be V(\phi) = V_1\left({V_0\over V_1} - {1\over \phi}\right) = {1.77 V_1} \approx 0.96\times 10^{-10} M_P^4,\ee
implying
\be V_1 = 10^{-10}~M_P^5, ~\mathrm{and}~ V_0 \approx 2\times 10^{-10}~M_P^4.\ee
In this manner all the parameters are fixed once and forever. Finally, we need to check if the model gracefully exits from inflation. In view of the form of the potential, $V(\phi) = V_0 - {V_1\over \phi_f}$, \eqref{2.33b} may be expressed as,
\be \label{ge1} {3H^2\over V_1}-{\gamma H^4\over V_1}={\dot\phi^2\over 2 V_1} + \left({V_0\over V_1}-{1\over\phi}\right).\ee
During inflation $\gamma H^4$ and $V_1$ are of the same order of magnitude, while Hubble parameter varies slowly. However, at the end, Hubble rate usually decreases sharply, and $\gamma H^4$ falls much below $V_1$. Hence $\gamma H^4 \over V_1$ falls of faster than $3H^2\over V_1$ and as a result, $\gamma H^4 \over V_1$ may be neglected without loss of generality. Hence, the above equation reads as,

\be \label{ge2} 6H^2=\dot\phi^2 -{2V_1\over\phi} + 2V_0.\ee
Clearly, $\phi$ exhibits oscillatory behaviour, provided $H$ does. In that case, $H$ should not decrease much even at the end of inflation, and the second term ${\gamma H^4\over V_1}$ of \eqref{ge1} cannot be neglected. However, the same result (oscillatory behaviour of $\phi$ as well as $H$) emerges keeping the second term as well. Thus, it is possible that the scalar field oscillates many times over a Hubble time, driving a matter-dominated era at the end of inflation. In the process, graceful exit from inflation is administered. Once scalar field is used up mostly by creating particles and reheating the universe, oscillation halts and a radiation dominated era initiates. The Hubble parameter $H$ then starts evolving differently. In a nut-shell, the model considered here perfectly fits with the evolution of the very early universe.\\

\subsection{Case-II}

In this subsection we consider the potential in the form: $V(\phi) = V_0 - V_1e^{-b\phi}$, so that $V' = b V_1e^{-b\phi}, ~~\mathrm{and},~~ V'' = -{V_1 b^2e^{-b\phi}}$. Fixing, $\phi_i = 6 M_P,~2\alpha = 1 M_p^2,~{V_0\over V_1} = 0.867, ~ \gamma V_1 = \sqrt 5 M_P^4 ~\mathrm{and}~ {b=0.45 M_P^{-1}}$, we compute the slow roll parameters, as:

\be \eta = -{b^2 e^{-b\phi}\over \left({V_0\over V_1}-e^{-b\phi}\right)}=-0.0170158\ee
\be \epsilon = \frac{\gamma^2V_1^2\left(b e^{-b\phi}\right)}{12\sqrt{2.25-{\gamma V_1\left({V_0\over V_1}-{e^{-b\phi}}\right)}}\left[1.5-\sqrt{2.25-{\gamma V_1\left({V_0\over V_1} -{e^{-b\phi}}\right)}}\right]^2}=0.000832993\ee
Therefore,

\be r = 16 \epsilon =0.0133279,~~~n_s = 1 - 6\epsilon + 2\eta =0.9609,~~~ \mathrm{and}~~~n_t = 0.0017,\ee
Further, at the value $\phi_f = 0.94 M_P$, $\epsilon = 1$, and slow rollover halts. The number of e-folds is found from the relation,

\be\begin{split} N =& {9\alpha\over \gamma}\int_{\phi_f}^{\phi_i} {d\phi\over V'} - {3\over \gamma}\int_{\phi_f}^{\phi_i}{\sqrt{9\alpha^2-\gamma V}\over V'}d\phi \end{split}=60\ee
Clearly, the fit is again excellent. In Table-2, we therefore organize a data set for the inflationary parameters, varying $\phi_i$ between $6 M_P\le \phi_i \le 6.3 M_P$, so that the spectral index of scalar perturbation lie within experimental limit.\\

\begin{figure}
\begin{center}
   \begin{minipage}[h]{0.47\textwidth}
      \centering
      \begin{tabular}{|c|c|c|c|c|}
      \hline\hline
    $\phi_i$ in $M_P$ & $n_s$ & $r$ & $\mathrm {N}$\\
    \hline
    6.00 & 0.9609 & 0.0133 & 60\\
    \hline
    6.05 & 0.9620 & 0.0127 & 62 \\
    \hline
    6.10 & 0.9630 & 0.0121 & 64 \\
    \hline
    6.15 & 0.9640 & 0.0115 & 66 \\
    \hline
    6.20 & 0.9650 & 0.0110 & 68 \\
    \hline
    6.25 & 0.9659 & 0.0105 & 70 \\
    \hline
    6.30 & 0.9668 & 0.0100 & 72 \\
    \hline\hline
    \end{tabular}
      \captionof{table}{Data set for the inflationary parameters under the choice $F(T) =\alpha T + \beta T^2$, where $\alpha = {M_P^2\over 2}$, for the potential $V_0=V_0 - V_1e^{-b\phi}$, taking into account, ${V_0\over V_1} = 0.867, ~ 108 \beta V_1 =\gamma V_1 = \sqrt 5 M_P^4 ~\mathrm{and}~ {b=0.45 M_P^{-1}}.$ Slow rollover ends at $\phi_f = 0.94 M_P$.}
      \label{tab:table1}
   \end{minipage}%
 \end{center}
\end{figure}

Let us now find the energy scale of inflation using relation \eqref{2.34c}, from the fourth data of the above table, associated with $N = 66$ as,
\be \label{2.34g} {H_*}^2 = \frac{1.5 -\sqrt{2.25-\gamma V_1\left({V_0\over V_1}-{e^{-b\phi_i}}\right)}}{\gamma}=\frac{0.8278}{\gamma },~~\mathrm{or} ~~H_* = \sqrt {0.8278\over \gamma}.\ee
where we have substituted $\alpha = {1\over 2} M_P^2$, ${V_0\over V_1} = 0.867$, $b = 0.45 M_P^{-1}$, $\phi_i = 6.15 M_P$ and $\gamma V_1 = \sqrt{5} M_P^4$. Note that, $\gamma$ remains arbitrary as before. Now, using the relation for single field inflation corresponding to GTR \cite{83} we find

\be \label{2.34h} H_* = 8\times 10^{13}\sqrt{r\over 0.2}~GeV \approx 7.89\times10^{-6} M_P,\ee
where, $r = 0.0115$ corresponding to the data associated with $N = 66$ of Table-2. Thus in order that the scale of inflation \eqref{2.34g} matches with the single field scale of inflation, $\gamma$ has to be of the order, $\gamma \approx 1.3\times 10^{10}$, i.e. $\beta \approx 10^8$. Note that the order of the dimensionless parameter $\beta$ remains unaltered from Case-I. Since individually each term has to be of the same order of magnitude, one can compare the two terms on the left hand side of \eqref{2.34a} to find,

\be 6\alpha H_*^2 \approx 1.9 \times 10^{-10} M_P^4,~~ \mathrm{while}~~108\beta H_*^4 = \gamma H_*^4 \approx 10^{-10}  M_P^4,\ee
provided $\gamma \approx 10^{10}$. This is essentially a consistency check. Note, that basically we had to fix only three parameters $\gamma V_1$, $V_0\over V_1$ and $b$ for a viable inflationary model. However, in view of the above consideration (sub-Planckian scale of inflation), the coupling parameter $\gamma = 108\beta$ also is fixed. In the same manner, one can also fix the numerical value of $V_1$ and hence $V_0$. Since the potential $V(\phi)$ appearing in \eqref{2.34a}, must also be of the same order of magnitude, so

\be V(\phi) = V_1\left({V_0\over V_1} -e^{-b\phi} \right) = 0.804 V_1 \approx 1.39\times 10^{-10},\ee
which implies

\be V_1 = 10^{-10}~M_P^5,~ \mathrm{and~hence},~ V_0 \approx 0.867\times 10^{-10}~M_P^4.\ee

\noindent
Now, in view of the potential $V(\phi) = V_0 - V_1e^{-b\phi}$, equation \eqref{2.33b} may be expressed as,
\be {3H^2\over V_1}-{\gamma H^4\over V_1}={\dot\phi^2\over 2 V_1} + \left({V_0\over V_1}-{e^{-b\phi}}\right)\ee
As before (Case-I), here again one might consider oscillatory behaviour of the scalar field $\phi$, provided $H$ oscillates as well at the end of slow rollover. In the process, graceful exit from inflation may be administered.

\section{Analytical solution in the radiation dominated era.}

The model under consideration unifies an early scalar-driven inflation with the late-stage of Hubble-parameter-driven accelerating universe, as already demonstrated in \cite{Inf16}, taking into account almost an identical form of $F(T)$. The inflationary parameters fit with latest released data sets from Planck \cite{Planck1, Planck2} with excellence. Inflation occurred at sub-Planckian scale and the model admits graceful exit from inflation too. In the present pressure-less dust era, the model might even cross the phantom divide line $(\Omega_\Lambda = -1)$, which is not excluded by observation. Such a wonderful outcome from the well-simplified model $(F(T) = \alpha T + \beta T^2)$, clearly motivates to study the later stage of cosmological evolution, when the oscillatory nature of the scalar field drives a matter dominated era, particularly the radiation dominated era to be precise. For this purpose, let us combine the first two field equations \eqref{2.33a} and \eqref{2.33b} to find:

\be \label{3.2} {4\over 3}\gamma H^2 \dot H - 4\alpha \dot H = \dot\phi^2 + (\rho + p).\ee
Additionally, we have the Bianchi identity, viz.

\be\label{BI} \dot \rho + 3 H(\rho + p) = 0.\ee
In the radiation dominated era $(p = {1\over 3}\rho)$, equation \eqref{BI} leads to $\rho = {\rho_{r0}\over a^4}$, where the constant $\rho_{r0}$ stands for the present value of the radiation density. Let us now seek power law solution of the scale factor in the form $a = a_0 t^n$, so that $n < 1$, to ensure decelerated expansion required for structure formation in the radiation dominated era. As a result, \eqref{2.33b} takes the form:

\be\label{3.3} {6\alpha n^2\over t^2} - {\gamma n^4 \over t^4} = {1\over 2}\dot\phi^2 + V(\phi) + {\rho_{r0}\over a_0^4 t^{4n}},\ee
due to Bianchi identity. It is important to mention that, if the scalar field is completely used up in the process of particle creation and the potential turns out to be a bare cosmological constant, then the above equation \eqref{3.3} is not satisfied due to the presence of three different powers of $(t)$ appearing in the denominator. Therefore, we express \eqref{3.3} in the following manner,

\be\label{3.4}{1\over 2}\dot \phi^2 + V(\phi) = {6\alpha n^2\over t^2}- {\gamma n^4 \over t^4} -{\rho_{r0}\over a_0^4 t^{4n}},\ee
which under time differentiation takes the form,

\be\label{3.5} \dot\phi\ddot \phi + \dot\phi V'(\phi) = -{12\alpha n^2\over t^3} + {4\gamma n^4 \over t^5} + {4n\rho_{r0}\over a_0^4 t^{4n+1}}.\ee
In view of \eqref{2.33c} and \eqref{3.5}, we therefore find,

\be\label{3.6} 3H\dot\phi^2 = {3n\dot\phi^2\over t} = {12\alpha n^2\over t^3} - {4\gamma n^4 \over t^5} - {4n\rho_{r0}\over a_0^4 t^{4n+1}}.\ee
Thus, we have,

\be\label{3.7} \phi = \phi_0 + \int\left[{4\alpha n\over t^2} - {4\gamma n^3 \over 3 t^4} - {4\rho_{r0}\over 3a_0^4 t^{4n}}\right]^{1\over 2} dt.\ee
If we now seek a typical Friedmann-like solution $a \propto \sqrt t$, i.e. $n = {1\over 2}$, then the above equation may be integrated to yield,

\be\label{3.7a} \phi = \phi_0 + c\ln\left[{c\left(ct + \sqrt{c^2 t^2-d^2}\right)}\right] - {\sqrt{c^2 t^2-d^2}\over t},\ee
where, $c^2 = 2\big(\alpha - {2\rho_{r0}\over 3a_0^4}\big)$ and $d^2 = {\gamma\over 6}$. Hence, in view of \eqref{3.4}, the potential may be expressed as
\be\label{pot} V(t) = \frac {\alpha}{2t^2}+\frac{\gamma}{48t^4}-\frac{\rho_{r0}}{3a^4_{0}t^2}.\ee
However, it is impossible to express the potential in terms of $\phi$. Nonetheless, it is apparent that the potential is in no way similar to either of the two \eqref{Vphi} chosen to study inflation. For example, as a special case if we choose $c = 0$, equivalently, $\rho_{r0} = {3\over 2}\alpha a_0^4$, then \eqref{3.7a} reduces to

\be \label{3.8} \phi =\phi_0 -{\sqrt{-d^2}\over t}  = \phi_0-\left(\sqrt{-{\gamma \over 6}}\right) t^{-1}.\ee
Clearly, we encounter a contradiction, since now real solution is admissible only if $\gamma < 0$, while we found $\gamma \thickapprox 10^{10}$ for viable inflation. In any case, if we chose $\gamma = -{\gamma_0}^2$, so that, $\phi = \phi_0 - \left(\sqrt{\gamma_0^2\over 6}\right) {1\over t}$, then we can use \eqref{3.4} to find

\be\label{3.9} V = {\gamma_0^2\over 48 t^4}, \ee
resulting in the following quartic form of $V = V(\phi)$ viz.,

\be\label{3.10} V(\phi) = {3\over 4\gamma_0^2} (\phi - \phi_0)^4.\ee
Thus, a viable Friedmann-like radiation dominated era requires a quartic potential, with a negative coupling parameter $\beta = \frac{\gamma}{108}$. Such a form of quartic potential is devoid of a flat region, and hence slow rollover is not admissible. Hence, despite the fact that unification of early inflation with late-time cosmic acceleration has been achieved, a viable radiation dominated era in the middle, remains obscure with the same form of potential.

\section{Concluding remarks.}

It is important to mention that $F(R)$ theory of gravity can unify early inflation with late-time accelerated expansion from purely geometric consideration, with a form like $F(R) = \alpha R + \beta R^m + \gamma R^{-n}$ where, $m > 0,~n > 0$ \cite{1.80}. However, cosmic evolution in the radiation dominated era is not at par with Friedmann-like evolution, unless one neglects $R^m$ term, which is not logical at that epoch. On the contrary, a particular form of $F(R) = \alpha R + \beta R^2 + \gamma R^{3\over 2}$ was found to encompass cosmic evolution right from the radiation era till date, in a continuous manner \cite{1.81}, while the presence of $R^2$ term can generate inflation in the very early universe \cite{1.82}. Thus, unless we obtain anything better, there is no point in considering yet another theory, such as `Teleparallel gravity'. It is noticeable that arbitrary form of $F(R)$ doesn't work. This is the reason for finding a suitable form of $F(R)$ over years, following either Noether symmetry analysis or reconstruction program.\\

Likewise, in order to study the cosmological consequence of the so-called Teleparallel gravity, a particular form of $F(T)$ is required. Usually, either Noether symmetry is imposed or reconstruction program is carried out to find such a form. Although, a host of $F(T)$ has been explored so far, it is learnt that not a single Noether conserved current associated with the available forms of $F(T)$ satisfy the $(^0_0)$ equation of Einstein \cite{NM}. However, reconstruction program could find some of these, the simplest being $F(T) = \alpha T + \beta T^2$. Note that $F(T) = \alpha T$, is simply GTR, apart from a total derivative term. Thus higher powers of $T$ is responsible for late-time cosmic acceleration. Hence, a scalar field (inflaton or Higgs) driven inflation has been studied extensively over last decade, taking into account some of these forms of $F(T)$. Inflationary parameters were matched with `the then' released Planck's data sets, graceful exit from inflation has been assured, and unification with late-time cosmic acceleration was found in some cases. It is therefore required primarily to match inflationary parameters with the latest released Planck's data set, and thereafter study the evolution in the radiation-dominated era, which initiated soon after graceful exit, as the universe enters hot big-bang era. This was our present motivation.\\

Since vacuum de-Sitter solution is admissible for arbitrary form of $F(T)$, so we choose the simplest form $F(T) = \alpha T + \beta T^2$, which envisages late-time cosmic acceleration assuring crossing of the phantom divide line as well. We then study slow roll inflation, being driven by a scalar field choosing two different types of potentials having flat sections, so that slow roll is applicable. The inflationary parameters viz. the tensor to scalar ratio $(r)$ and the spectral index of scalar perturbation $(n_s)$ exhibit wonderful fit with the recent released data from Planck's collaborators, keeping the number of e-folds around $N = 60$. In the process, unification is achieved. However, the potentials responsible for slow-roll does not ensure a decelerated Friedmann-like expansion in the radiation dominated era. On the contrary, the quartic potential required for this purpose, is devoid of a flat section, and hence slow-roll is not admissible.\\

Of-course, slow roll inflation with quartic potential may be studied, but in more involved theories of gravity, e.g. in non-minimally coupled scalar-tensor theories \cite{sami, beh, dalia} or in higher order theories \cite{ranajit1, ranajit2, subhra}, in which an effective potential with a flat section comes into play. It is important to mention that $F(T)$ theories have also given rise to some disputes about an intriguing and essential feature: for example, the action of the theory is not invariant under local Lorentz transformations of the tetrad \cite{1.39, 1.50, 1.72, 1.72a}, and also it suffers from the unresolved pathology of `Branched Hamiltonian' that we shall discuss in future.

\end{document}